\documentclass[3p,times]{elsarticle}

\usepackage{amsmath,bm,amsfonts,amssymb,bbm}
\usepackage{color}

\usepackage{amssymb}

\usepackage[figuresright]{rotating}
\usepackage{wasysym}
\begin{document}

\begin{frontmatter}
\title{Sign Problem in Quantum Monte Carlo Simulation}

\author[label1,label2]{Gaopei Pan}
\author[label3]{Zi Yang Meng}
\address[label1]{Beijing National Laboratory for Condensed Matter Physics and Institute of Physics, Chinese Academy of Sciences, Beijing 100190, China}
\address[label2]{School of Physical Sciences, University of Chinese Academy of Sciences, Beijing 100190, China}
\address[label3]{Department of Physics and HKU-UCAS Joint Institute of Theoretical and Computational Physics, The University of Hong Kong, Pokfulam Road, Hong Kong SAR, China}

\begin{abstract}
Sign problem in quantum Monte Carlo (QMC) simulation appears to be an extremely hard yet interesting problem. In this article, we present a pedagogical overview on the origin of the sign problem in various quantum Monte Carlo simulation techniques, ranging from the world-line and stochastic series expansion Monte Carlo for boson and spin systems to the determinant and momentum-space quantum Monte Carlo for interacting fermions. We point out the basis dependency of the sign problem and summarize the progresses to cure, ease and even make use of the sign problem over the years, such as symmetry analysis of the underlying Hamiltonian, basis optimization in writting down the partition functions and many others. Moreover, we state that although traditional lore saying that in case of sign problem, the average sign in QMC simulation approaches zero exponentially fast with the space-time volume of the configurational space, there are recent breakthroughs showing this is not always the case and based on the properties of the partition function for finite size systems, one could distinguish when the average sign has the usual exponential scaling and when it is bestowed with an algebraic scaling at the low temperature limit. Fermionic QMC simulations with such algebraic sign problems have been successfully carried out for extended Hubbard-type and quantum Moir\'e lattice models. 
\end{abstract}


\end{frontmatter}

\section{Introduction}

Quantum Monte Carlo (QMC) method is an extremely powerful and unbiased method for studying strongly correlated systems, especially in condensed matter, high energy and quantum material research~\cite{Foulkes2001,Carlson2015,Assaad2008,Sandvik2010,XYXu2019}, where the quantum many-body lattice models are in general analytically intractable. In QMC simulation, the partition function of the interacting problem is cast into a sum (or integral) over configurations in a chosen basis, and the important sampling scheme could in principle cover the exponentially large configurational space in polynomial time.

However, such practice is oftentimes hindered by the infamous "Sign Problem"~\cite{Sugar1990exp}. Sign problem means the weights of configurations in the QMC simulation become negative or even complex, and therefore the configurational weights in the Monte Carlo process cannot be further interpreted as classical probabilities, and the minus or even complex sign will rendered the QMC simulation with exponential complexity for obtaining the data with the same quality compared with the simulations without sign problem. It has been proved that if we could find a method to solve nondeterministic polynomial (NP) hard problem efficiently, the method can be used to solve the QMC simulations with the sign problem~\cite{NP-hard}. It's important to note that this doesn't mean the sign problem is considered to be NP-hard \cite{kaul2013review}, nor is it inconsistent with performing polynomial-time QMC simulation of specific cases that have negative or complex sign.

The origin of the sign problem has no universal explanation, but it is believed to come from the quantum mechanic properties of the constituent particles in the many-body systems, given them fermions, bosons or spins. Frustrated spin systems often have a sign problem because the off-diagonal operators in the Hamiltonian usually bring a negative amplitude~\cite{Frustrated,XXZ} and the number of off-diagonal operators can be odd for a particular configuration. While in fermionic systems, negative weights usually arise from the Pauli exclusion principle~\cite{NP-hard} or different Hubbard-Stratonovich decoupling~\cite{Hirsch1985} scheme (which means decoupling the quartic fermion interaction term into certian quadratic basis). At the same time, there are also proposals that the negative sign of a configuration is a topological invariant, which is an imaginary time counterpart of the Aharonov-Anandan phase and can be reduced to a Berry phase in the adiabatic limit~\cite{Topological_Origin}.  And the sign problem has even been linked to quantum phase transitions~\cite{2021QPT,Tarat2022}.  Moreover, it was shown recently that some interacting models may have intrinsic sign problem, which cannot be cured~\cite{hastings2016quantum,ringel2017quantized,Intrinsic1,Intrinsic2}. In general, an unified description of the sign problem is still missing. These different situation where the sign problem occur in QMC simulations will be discussed in a case based manner in Secs.~\ref{sec:sec2} and ~\ref{sec:sec3} of this article.

Fortunately, there are many cases of quantum lattice models where there are no or weak (polynomial instead of exponential scaling) sign problems and Monte Carlo calculations can be carried out with polynomial complexity. In frustrated magnet lattice models, there is no sign problem for some bilayer models \cite{Frustrated_bilayer,Frustrated_dimer,Frustrated_ladders,bilayer_again,bilayer_b} within certain parameters. Inspired by the absence of sign problems in the above models, one can choose those models as sign-problem-free reference system and extend the simulation basis from spin to singlet-triplet or even plaquette bases, and it has been shown the QMC simulations can be carried out to the parameter space for few important 2D frustrated lattice models (such as the Shastry-Sutherland antiferromagnetic spin model) where there were sign problem in the original spin basis~\cite{SSmodel,SSmodel_basis}. Interacting fermion models (mainly Hubbard-type) at certain basis with certain symmetries, such as anti-unitary symmetry, have been proved to be sign problem free~\cite{2005CongjunWu,Lang1993,Koonin1997,Hands2000}. Flat band quantum Mori\'e lattice models with $C_2 T$ + $C_2 P$ symmetries can also ensure that there are no sign problems~\cite{XuZhang2021,JYLee2021,ouyang2021projection,zhang2021sign}.  There also exist cases where one can use Majorana representation to avoid the sign problem, to split the fermionic operator into Majorana fermions and reuse the anti-unitary symmetry~\cite{HongYao1,HongYao2,HongYao3,2016TaoXiang,2015LeiWang}. In addition, basis transformation~\cite{2015BasisChange,2021BasisChange2,2020BasisChange3,2020BasisChange4,2020BasisChange5,2019BasisChange6,2020BasisChange7,Rossi2017determinant,Rossi2017polynomial,DEmidio2020}, Lefschetz thimbles~\cite{Lef2,Lef3}, meron cluster~\cite{Chandrasekharan1999}, Fermi bags~\cite{Huffman2014}, split-orthogonal group~\cite{2015LeiWang}, Majorana positivity~\cite{2016TaoXiang}, semigroup approach~\cite{Wei2017} and pesudo-unitary group~\cite{Xu2019}, automatic differentiation~\cite{wan2020mitigating}, machine learning techniques ~\cite{broecker2017machine,2021Wynen} and adiabatic method~\cite{adiabatic2021}, or through some systematic expansion\cite{Perturbative} can also solve or ease some sign problems. These cases will be discussed in Secs.~\ref{sec:sec4} and ~\ref{sec:sec5} of this article.

If there is sign problem in the QMC simulation, the usual simulation approach is to reweight: one can take the magnitude of the original weight (or magnitude of the real part) as the new weight, and add the sign into the sampling process of physical observables and divide the average sign $\langle s \rangle$ . This approach does not solve the sign problem, but only cast it into a different form. The average sign over the Monte Carlo sampling will appear in the denominator, so the average sign will affect how long it will take to obtain meaningful expectation values with controlled errorbars in the simulation. Usually $\langle s \rangle \propto e^{-N \beta \Delta f}$ where $\Delta f$ is the free energy difference of the two systems before and after the Monte Carlo update. This means sign often decay exponentially. And there are a lot of research which support viewpoint. No matter what lattice (hypercubic, ladder, depleted square, Lieb,
honeycomb, kagome, and triangular lattices) or parameter (temperature, interaction strength, and density), the sign usually decay exponentially. There is a good summary \cite{Geometry} of the sign datas for different geometry and parameters.

  But Recent developments have changed this viewpoint, there are cases in Hubbard model at certain filling the average sign is not exponential decay~\cite{tarat2021,ibarra2021universal} and we have found that it is possible to prove the average sign of quantum many-body lattice models accquire algebraic sign structure~\cite{ouyang2021projection,zhang2021sign}, if their finite size partition functions satisfy the {\it{Sign bound theory}}. The details on the theory and its application on the systems with extended or long-range interaction such as the extended Hubbard~\cite{ouyang2021projection,YDLiao2022} and quantum Moir\'e lattice models~\cite{XYXu2018Kekule,LiaoCPB2021,YDLiaoPRX2021,XuZhang2021,GaopeiPanValley2021,zhang2021sign,XuZhang2021SC}, are given in Sec.~\ref{sec:sec5.3} of this article.  

\section{Configurational weight in classical Monte Carlo simulation}
\label{sec:sec2}
Before presenting the details of the sign problem, we first discuss classical Monte Carlo simulation briefly such that the origin of the sign problem manifests naturally. 

In classical Monte Carlo, the partition function at temperature $T$ is written as follows:
\begin{equation}
	Z=\operatorname{Tr}[e^ {-\beta H}]=\sum_{\mathcal{C} \in \Omega} e^{-\beta E(\mathcal{C})}=\sum_{\mathcal{C} \in \Omega} W(\mathcal{C})
\end{equation}
where $\beta=\frac{1}{k_B T}$, $k_B$ is the Boltzmann constant, and  $\mathcal{C}$ stands for the configurations in the configurational space $\Omega$, which is exponetially large in system size, such as $2^{L^d}$ and $4^{L^d}$ for $d$ dimensional spin or fermion systems with linear dimension $L$. Once the partition function is written as a sum of Boltzmann weights $e^{-\beta E(\mathcal{C})}$, it's clear that in the classical Monte Carlo the $W(\mathcal{C})$ corresponding to different configurations are always positive and real numbers. So one can easily carry Monte Carlo updates in Markov chain by calculating the ratio of weights of different configurations.

\begin{figure}
	\centering
	\includegraphics[width=0.7\linewidth]{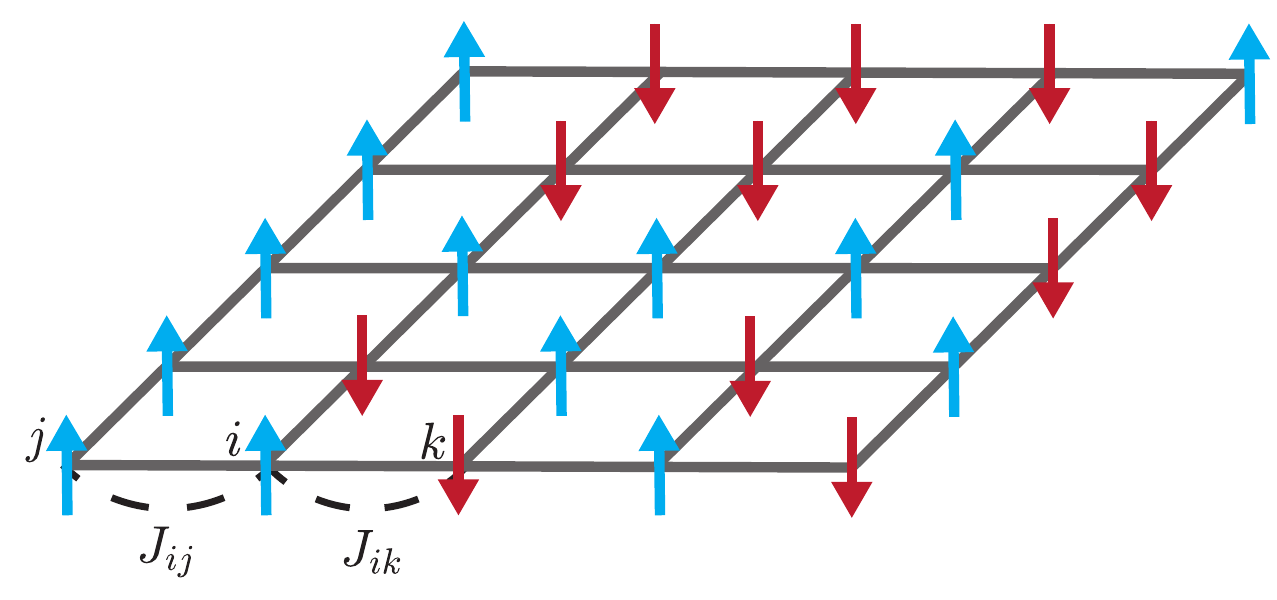}
	\caption{Classical Ising configuration of the example $E(\mathcal{C})=\sum_{i,j}E_{i,j}$ and $E_{i,j}=J_{i,j}$ and $E_{i,k}=-J_{i,k}$.}
	\label{fig:fig1}
\end{figure}

For example, as shown in Fig.~\ref{fig:fig1}, the Hamiltonian of the classical 2d Ising model is: $H= \sum_{ i, j }J_{i,j} s_{i} s_{j}$, where $s_{i}$  denotes the state of the $i$-th spin which takes the value of $\pm 1$ (up or down), $J_{i,j}$ is the interaction strength. For the configuration  in Fig.~\ref{fig:fig1}, the corresponding weight is $e^{-\beta \sum_{i,j} E_{i,j}(\mathcal{C})}$.

We know the expectation values of observable $O$ is:
\begin{equation}
	\langle O\rangle=\frac{1}{Z} \operatorname{Tr}[O e^ {-\beta H}]=\frac{1}{Z} \sum_{\mathcal{C} \in \Omega} O(\mathcal{C}) W(\mathcal{C}).
\end{equation}
For the classical case, since $W(\mathcal{C})$ is a real number and $W(\mathcal{C}) \geqslant  0$, one can carry out the Monte Carlo and obtain $N_b$ configurations $\left\{\mathcal{C}_{i}\right\}$ from $\Omega$, according to the distribution $p\left(\mathcal{C}_{i}\right)=W\left(\mathcal{C}_{i}\right) / Z$. Here $N_b$ is number of sampling bins, and then the average value of the samples is the expectation value of the physical quantity $O$:

\begin{equation}
	\langle O\rangle \approx \bar{O}=\frac{1}{N_b} \sum_{i=1}^{N_b} O\left(\mathcal{C}_{i}\right).
\end{equation}

The statistical error corresponding to such an estimate of the mean value is~\cite{geyer2011introduction}:
\begin{equation}
	\Delta O=\sqrt{\frac{\operatorname{Var} (O)\;2 \tau_{O}^{\text {int }} }{N_b}},
\end{equation}
where $\operatorname{Var} (O)$ is the variance of $O$ and the integrated autocorrelation time $\tau_{O}^{\text {int }}$ is a measure of the autocorrelations of the sequence $\left\{O\left(\mathcal{C}_{i}\right)\right\}$
\begin{equation}
	\tau_{O}^{\text {int }}=\frac{1}{2}+\sum_{t=1}^{\infty} A_{O}(t),
\end{equation}
and the autocorrelation function $A_{O}(t)$ for a quantity $O$ is defined as:
\begin{equation}
	A_{O}(t)=\frac{\langle O(\mathcal{C}_{i+t}) \,O(\mathcal{C}_{i})\rangle-\langle O\rangle^{2}}{\left\langle O^{2}\right\rangle-\langle O\rangle^{2}}.
\end{equation}

In principle, since in classical Monte Carlo weights are always real and positive and the statistical errors are inversely proportional to the number of bins $\Delta O \propto \frac{1}{\sqrt{N_b}}$. One can carry out Monte Carlo sampling in polynomial time to achieve controlled expectation values (critical slowing down is not taken into account).  

\section{Configurational weight in quantum Monte Carlo simulation}
\label{sec:sec3}
However when we consider quantum Monte Carlo (QMC) simulation, the weight may become a negative or even a complex number and cause the sign problem in the sampling. In QMC, quantum many-body problems in $d$ dimension are usually mapped onto a classical system in $\mathrm{d}+z$ dimension, where $z$ is the dynamic exponent of the problem and usually taken to be 1 in practically simulations~\cite{d+1map}. Depending on the quantum lattice models, there are many different QMC algorithms, such as world-line Monte Carlo~\cite{d+1map,loop,worm,Worldline_fermion}, stochastic series expansion (SSE)~\cite{sandvik1991quantum,sandvik1992generalization}, determinant quantum Monte Carlo (DQMC)~\cite{Hirsch1985,DQMC}, momentum space quantum Monte Carlo~\cite{XuZhang2021,GaopeiPanValley2021,XiDai2022} and so on. The form of the weights in these methods and the origin of sign problem are not the same and we will discuss them one by one.

\subsection{World-Line Monte Carlo}
The usual world-line approach~\cite{d+1map,loop,worm,de1985monte,Worldline_fermion,wl_boson} starts with discretizing the inverse temperature $\beta$, and employing the Suzuki-Trotter approximation~\cite{trotter1959product,suzuki1985general,fye1986new,fye1987calculation} to decompose the exponential of the Hamiltonian. A common world-line expansion of the partition function is: 
\begin{equation}
	\begin{gathered}
		Z=\operatorname{Tr} [e^{-\beta \hat{H}}]=\operatorname{Tr} [e^{-M \Delta \tau \hat{H}}]=\operatorname{Tr}\left[\left(e^{-\Delta \tau \hat{H}_{1}} e^{-\Delta \tau \hat{H}_{2}}\right)^{M}\right]+\mathcal{O}\left(\Delta \tau^{2}\right) \\
		=\sum_{i_{1}, \ldots, i_{2M} }\left\langle i_{1}\left|e^{-\Delta \tau \hat{H}_{1}}\right| i_{2 M}\right\rangle\left\langle i_{2 M}\left|e^{-\Delta \tau \hat{H}_{2}}\right| i_{2 M-1}\right\rangle \cdots \\
		\left\langle i_{3}\left|e^{-\Delta \tau \hat{H}_{1}}\right| i_{2}\right\rangle\left\langle i_{2}\left|e^{-\Delta \tau \hat{H}_{2}}\right| i_{1}\right\rangle
	\end{gathered}
\end{equation}
where $|i_l\rangle\langle i_l|$, $l\in[1,2M]$ are the complete basis of the quantum many-body problem and they are inserted between the imaginary time evolution operators. Separating $\hat{H}=\sum_{i} \hat{H}_{i}$ in such a way to facilitate the calculation of matrix elements make it easy to consider the sign problem. We can see that the matrix elements $\left\langle i\left|e^{-\Delta \tau \hat{H}_{k}}\right| j\right\rangle$ in the weights of a worldline configuration are not obviously positive or negative, therefore, we in general don't know whether the product of these matrix elements is a positive real number.  

\begin{figure}[!ht]
	\centering
	\includegraphics[width=0.8\linewidth]{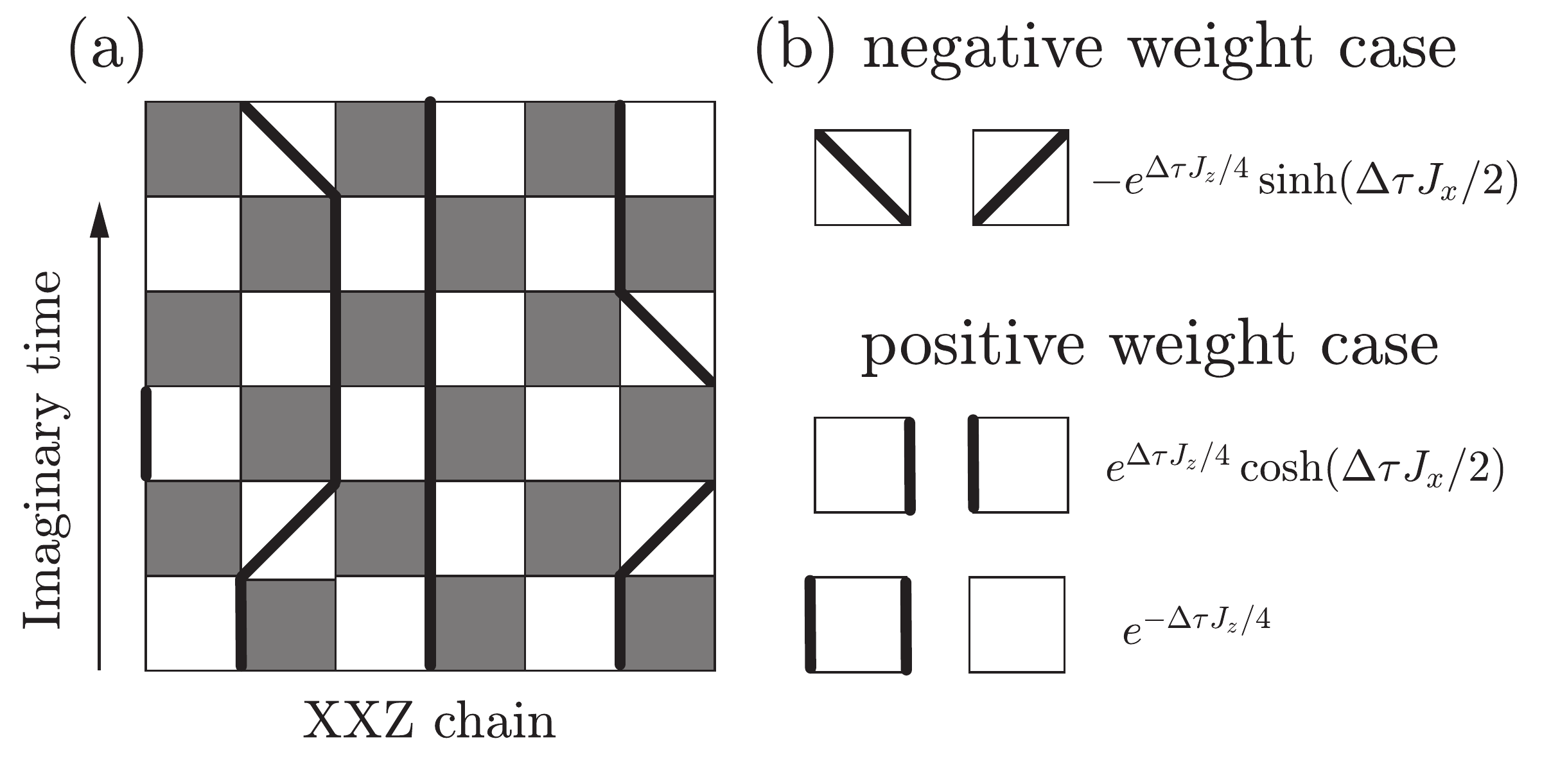}
	\caption{ World line configuration for the XXZ model. (a). The bold lines correspond to the imaginary time evolution of spin up states, and the unpassed points correspond to states with spin down. (b). Corresponding weights of matrix elements in different cases. We see that the negative weight case corresponds to the spin flip operation.}
	\label{fig:fig2}
\end{figure}

For example, for spin-1/2 XXZ model in a chain, with Hamiltonian:
\begin{equation}
	\hat{H}=J_{x} \sum_{i}\left(\hat{S}_{i}^{x} \hat{S}_{i+1}^{x}+\hat{S}_{i}^{y} \hat{S}_{i+1}^{y}\right)+J_{z} \sum_{i} \hat{S}_{i}^{z} \hat{S}_{i+1}^{z}.
\end{equation}
We can split it in such a way:
\begin{equation}
	\hat{H}=\underbrace{\sum_{n} \hat{H}^{(2 n+1)}}_{\hat{H}_{1}}+\underbrace{\sum_{n} \hat{H}^{(2 n+2)}}_{\hat{H}_{2}}
\end{equation}
where $\hat{H}^{(i)}=J_{x}\left(\hat{S}_{i}^{x} \hat{S}_{i+1}^{x}+\hat{S}_{i}^{y} \hat{S}_{i+1}^{y}\right)+J_{z} \hat{S}_{i}^{z} \hat{S}_{i+1}^{z}$. By introducing spin flip operators $\hat{S}_{j}^{\pm}=\hat{S}_{j}^{x} \pm i \hat{S}_{j}^{y}$, we can write configurations in the spin representation. An example of QMC configuration and values of matrix elements can be seen in  Fig~ \ref{fig:fig2}. In this case, although the corresponding weight of spin flip is negative, the periodic boundary condition guarantees that the number of spin flips must be even, so the total weight must be positive.  But for other cases, there may not be similar conditions to ensure that the weight is positive.  

For continuous-time world-lines, the expansion may vary, but the form of the weights is broadly similar:

If $\hat{H}=\hat{H}_0+\hat{V}$, and $\hat{V}$ is a perturbation~\cite{prokof1998exact}. We define $e^{-\tau \mathcal{\hat{H}}_{0}} \hat{U}(\tau)=e^{-\tau\left(\mathcal{\hat{H}}_{0}+\hat{V}\right)}$ and $\hat{V}(\tau)=e^{\tau \mathcal{\hat{H}}_{0}} \hat{V} e^{-\tau \mathcal{\hat{H}}_{0}}$. Then 
\begin{equation}
	\begin{aligned}
		Z=& \operatorname{Tr}\left[e^{-\beta \mathcal{\hat{H}}_{0}} \hat{U}(\beta)\right] \\
		=& \operatorname{Tr}\left[e^{-\beta \mathcal{\hat{H}}_{0}}\left(1-\int_{0}^{\beta} d \tau_{1} \hat{V}\left(\tau_{1}\right)+\int_{0}^{\beta} d \tau_{2} \int_{0}^{\tau_{2}} d \tau_{1} \hat{V}\left(\tau_{2}\right) \hat{V}\left(\tau_{1}\right)-\cdots\right)\right] \\
		=&\operatorname{Tr}\left[\sum_{n=0}^{\infty}(-1)^{n} \int_{0}^{\beta} d \tau_{n} \int_{0}^{\tau_{n}} d \tau_{n-1} \cdots \int_{0}^{\tau_{2}} d \tau_{1} e^{-\left(\beta-\tau_{n}\right) \mathcal{\hat{H}}_{0}} \hat{V} e^{-\left(\tau_{n}-\tau_{n-1}\right) \mathcal{\hat{H}}_{0}}\times \hat{V} \cdots \hat{V} e^{-\tau_{1} \mathcal{\hat{H}}_{0}}\right] \\
		=& \sum_{n=0}^{\infty}(-1)^{n} \sum_{\phi_{n}, \cdots, \phi_{1}}^{\beta} \int_{0}^{\tau_{2}} d \tau_{n} \int_{0}^{\tau_{n}} d \tau_{n-1} \cdots \int_{0}^{\tau_{2}} d \tau_{1}\left\langle\phi_{1}\left|e^{-\left(\beta-\tau_{n}\right) \mathcal{\hat{H}}_{0}}\right| \phi_{1}\right\rangle\left\langle\phi_{1}|\hat{V}| \phi_{n}\right\rangle \\
		& \times\left\langle\phi_{n}\left|e^{-\left(\tau_{n}-\tau_{n-1}\right) \mathcal{\hat{H}}_{0}}\right| \phi_{n}\right\rangle\left\langle\phi_{n}|\hat{V}| \phi_{n-1}\right\rangle \cdots\left\langle\phi_{2}|\hat{V}| \phi_{1}\right\rangle\left\langle\phi_{1}\left|e^{-\tau_{1} \mathcal{\hat{H}}_{0}}\right| \phi_{1}\right\rangle
	\end{aligned}
\end{equation}
where $\hat{U}(\tau)=1-\int_{0}^{\tau} d \tau^{\prime} \hat{V}\left(\tau^{\prime}\right) \hat{U}\left(\tau^{\prime}\right)$.

\subsection{Stochastic Series Expansion}
\label{sec:sec3.2}
Stochastic series expansion (SSE) method~\cite{sandvik1991quantum,sandvik1992generalization} Taylor expands the partition function, and it can be shown that the expansion of finite size systems can be safely truncated,
\begin{equation}
	\begin{aligned}
		Z & =\operatorname{Tr}[ \exp^{-\beta \hat{H}}]=\sum_{n=0}^{\infty} \frac{(-\beta)^{n}}{n !} \operatorname{Tr}[ \hat{H}^{n}]  \\
		& =\sum_{n=0}^{\infty} \sum_{i_{1}, \ldots, i_{n}} \frac{(-\beta)^{n}}{n !}\left\langle i_{1}|\hat{H}| i_{2}\right\rangle\left\langle i_{2}|\hat{H}| i_{3}\right\rangle \cdots\left\langle i_{n}|\hat{H}| i_{1}\right\rangle \\
		& \equiv \sum_{n=0}^{\infty} \sum_{i_{1}, \ldots, i_{n}} p\left(i_{1}, \ldots, i_{n}\right) \equiv \sum_{\mathcal{C} \in \Omega} W(\mathcal{C})
	\end{aligned}
\end{equation}
which is similar to the case of the world-line, the final sign of the weight is also determined by the number of negative matrix elements multiplied, i.e., usually by the number of off-diagonal elements (on the spin $\hat{S}^z_i$ basis, the spin flip term $\hat{S}_{j}^{+}\hat{S}_{j+1}^{-}+\hat{S}_{j}^{-}\hat{S}_{j+1}^{+}$ is off-diagonal).  For bipartite lattices such as square or honeycomb with periodic boundary condition in space-time, it can be shown that the number of the off-diagonal operator is always even such that their product in the configurational weight will be positive, however, for non-bipartite lattices, such as triangle and kagome, the number of the off-diagonal operators can be odd and the SSE simulation will have sign problem. We note that there are recent developments in frustrated magnet lattice models, some bilayer models~\cite{Frustrated_bilayer,Frustrated_dimer,Frustrated_ladders,bilayer_again,bilayer_b} within certain parameters with cluster basis are sign-problem-free. Inspired by these works and considering the model that has the same ground state as the above model, one can extend the simulation basis from spin to singlet-triplet or even plaquette bases, and it has been shown the SSE-QMC simulations for few important 2D frustrated lattice model, such as the Shastry-Sutherland antiferromagnetic spin model, can be carried out to the parameter space where there were sign problem in the original spin basis~\cite{SSmodel,SSmodel_basis}.

\subsection{DQMC}
For interacting fermion models, one use different ways to write down the partition functions. In determinant quantum Monte Carlo (DQMC)~\cite{Hirsch1985,DQMC}, auxiliary fields are introduced to decouple interactions through Hubbard-Stratonovich transformation. Taking Hubbard model as an example, Suzuki-Trotter decomposition is still necessary: 
\begin{equation}
	Z=\operatorname{Tr}\left[e^{-\Delta \tau \hat{H}_{I}} e^{-\Delta \tau \hat{H}_{0}} e^{-\Delta \tau \hat{H}_{1}} e^{-\Delta \tau \hat{H}_{0}} \cdots e^{-\Delta \tau \hat{H}_{1}} e^{-\Delta \tau \hat{H}_{0}}\right] 
\end{equation}
where $\hat{H}_0=-t\sum_{\langle i,j \rangle}(\hat{c}_i^{\dagger}\hat{c}_j +h.c.)$ and $\hat{H}_{I}=U\sum_{i}\left(\hat{n}_{i \uparrow}-\frac{1}{2}\right)\left(\hat{n}_{i \downarrow}-\frac{1}{2}\right)$. $\langle i,j \rangle$ means nearest-neighbour hopping. Then Hubbard-Stratonovich transformation gives:
\begin{equation}
	\begin{aligned}
		e^{-\Delta \tau \hat{H}_{I}}&=\prod_{i} e^{-\Delta \tau U\left(\hat{n}_{i \uparrow}-\frac{1}{2}\right)\left(\hat{n}_{i \downarrow}-\frac{1}{2}\right)}  \\
		&=\prod_{i} \gamma \sum_{s_{i, l}=\pm 1} e^{\alpha s_{i, l}\left(\hat{n}_{i \uparrow}-\hat{n}_{i \downarrow}\right)}    \\	
		&=\gamma^{N} \sum_{s_{i, l}=\pm 1}\left(\prod_{i} e^{\alpha s_{i, l} \hat{n}_{i \uparrow}} \prod_{i} e^{-\alpha s_{i,l} \hat{n}_{i \downarrow}}\right)
	\end{aligned}
\end{equation}
where $\gamma=\frac{1}{2} e^{-\Delta \tau U / 4}, \cosh (\alpha)=e^{\Delta \tau U / 2}$. 

Since we have $\operatorname{Tr}\left[e^{-\sum_{i, j} \hat{c}_{i}^{\dagger} A_{i, j} \hat{c}_{j}} e^{-\sum_{i, j} \hat{c}_{i}^{\dagger} B_{i, j} \hat{c}_{j}}\right]=\operatorname{Det}\left(1+e^{-\mathbf{A}} e^{-\mathbf{B}}\right)$, then partition function is:
\begin{equation}
	\begin{aligned}
		Z&=\sum_{s_{i,l},=\pm 1} \gamma^{N L_{\tau}} \operatorname{Tr}_{F}\left\{\prod_{l=1}^{L_{\tau}}\left[\left(\prod_{i} e^{\alpha s_{i,l}, \hat{n}_{\uparrow} }\right)\left(e^{-\Delta \tau c_\uparrow^{\dagger} T c_{\uparrow}}\right)\left(\prod_{i} e^{-\alpha s_{i,l}, \hat{n}_{i,\downarrow}}\right)\left(e^{-\Delta \tau c^{\dagger}_\downarrow T c_{\downarrow}}\right)\right]\right\}\\
		&=\gamma^{N L_{\tau}} \sum_{\left\{s_{i, l}\right\}} \prod_{\sigma=\uparrow \downarrow} \operatorname{Det}\left[\mathbf{I}+\mathbf{B}^{\sigma}\left(L_{\tau} \Delta \tau,\left(L_{\tau}-1\right) \Delta \tau\right) \cdots \mathbf{B}^{\sigma}(\Delta \tau, 0)\right]
	\end{aligned}
\end{equation}
where 
\begin{equation}
	\begin{aligned}
		&\mathbf{B}^{\uparrow}\left(l_{2} \Delta \tau, l_{1} \Delta \tau\right)=\prod_{l=l_{1}+1}^{l_{2}} e^{\alpha \operatorname{Diag}\left(\vec{S}_{l}\right)} e^{-\Delta \tau T} \\
		&\mathbf{B}^{\downarrow}\left(l_{2} \Delta \tau, l_{1} \Delta \tau\right)=\prod_{l=l_{1}+1}^{l_{2}} e^{-\alpha \operatorname{Diag}\left(\vec{S}_{l}\right)} e^{-\Delta \tau T}
	\end{aligned}
	\label{eq:eq15}
\end{equation}

and $\operatorname{Diag}\left(\vec{S}_{l}\right)$ is a diagonal matrix with diagonal elements $\boldsymbol{s}_{i, l}$. For the space-time configuration $\{s_{i,l}\}$, the calculation of weight is converted to the calculation of the determinant of matrices. Without a prior knowledge such as symmetry analysis or energy spectrum~\cite{zhang2021sign}, the determinant of a matrix doesn't have to be a positive real number. It is worth noting that the Hubbard-Stratonovich transformation which introduce auxiliary field is also one of the causes of the sign problem. For example, using the world-line method, the one-dimensional Hubbard model has no sign problem\cite{Assaad2008,Worldline_fermion}, but for DQMC, usually there is a sign problem\cite{Geometry}.    

\begin{figure}[!ht]
	\centering
	\includegraphics[width=0.8\linewidth]{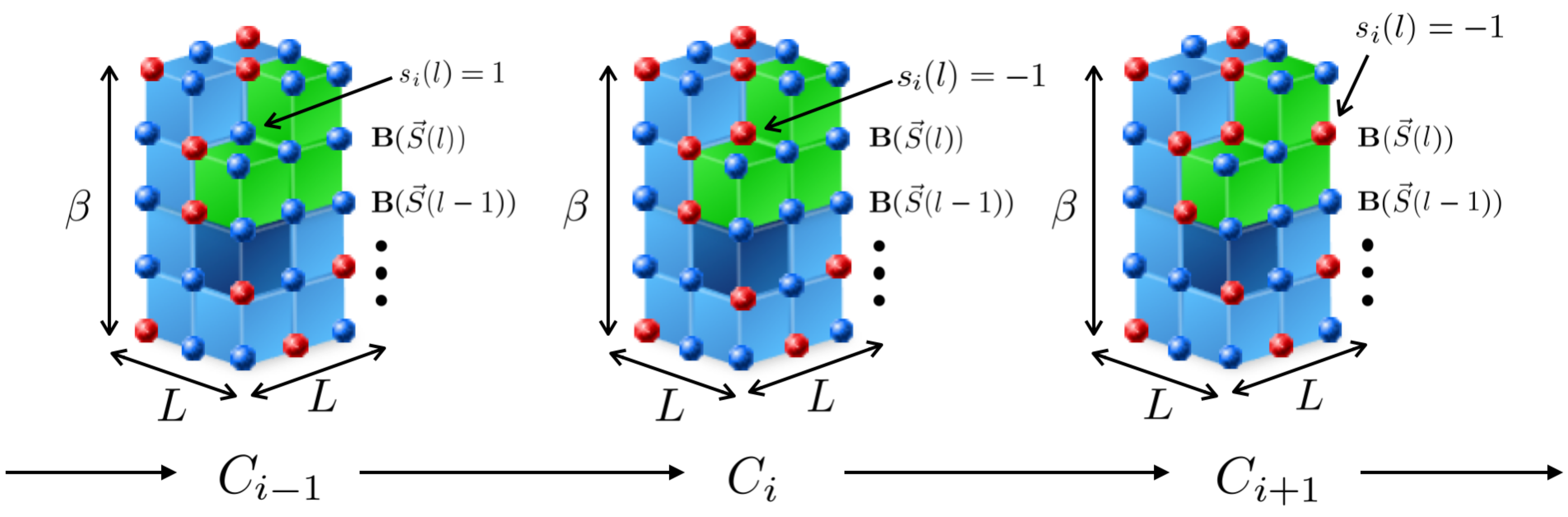}
	\caption{The space-time auxiliary field $\{s_{i,l}\}$ introduced by Hubbard-Stratonovich transformation  lives on the lattice of $d+1$ dimension (here $d=2$ with linear system size $L$), and we perform the calculation of Markov chain Monte Carlo by flipping the auxiliary field on the space-time lattice with the corresponding probability. Here the notation $\mathbf{B}(\vec{S}(l))$ corresponds to the matrix containing the auxiliary field of imaginary time layer $l$, shown in Eq.~\eqref{eq:eq15}.}
	\label{fig:DQMC}
\end{figure}

A Markov chain for the DQMC simulation for the 2d Hubbard model, with $L\times L\times \beta$ and the auxiliary fields for three consecutive configurations are given in Fig.~\ref{fig:DQMC}.

\section{What is the Sign Problem?}
\label{sec:sec4}
With the above preparation, we now discuss the real content of the sign problem. As the name implies, the simplest and most straightforward defination of the sign problem is that,  the weights (probabilities) in the Monte Carlo simulation  $p\left(\mathcal{C}_{i}\right)=W\left(\mathcal{C}_{i}\right) / Z$ is negative or becomes a complex number.  

For the real time evolution of a quantum mechanic state, this is obviously going to happen, because the time evolution operator is $e^{-i t H}$ and the corresponding configurational weight $W(\mathcal{C}) \sim \exp (i S(\mathcal{C}))$, here $S(\mathcal{C})$ is the action function of configuration $\mathcal{C}$, which is time integral of Lagrangian $L$: $S = \int d t \,L(x, \dot{x})$. It's not surprising that now the weight might be a complex number.  While if we consider the ensemble average at the thermal equilibrium state for classical statistical problems such as the Ising model in Sec.~\ref{sec:sec2}, configurational weight $W(\mathcal{C}) \sim \exp (-\beta H(\mathcal{C}))$, is real and positive.

When we discuss algorithms that do not suffer from the sign problem~\cite{NP-hard}, it usually means that the algorithm needs polynomial complexity to evaluate the thermal average $\langle O\rangle$, i.e. with the system size $N=L^d$ and inverse temperature $\beta=\frac{1}{k_B T}$, the computational time required to make the relative error $\Delta O /\langle O\rangle$ less than $\epsilon$ is called $t(\epsilon, N, \beta)$. If $t(\epsilon, N, \beta)$ grows polynomially with $N$ and $\beta$, i.e. there exist integers $n$ and $m$ and a constant $\kappa<\infty$ such that
\begin{equation}
	t(\epsilon, N, \beta)<\kappa \epsilon^{-2} N^{n} \beta^{m}.
\end{equation}
Such complexity is tractable with realistic Monte Carlo simulation in both human and machine time.

One thing needs to clarify is that sign problem doesn't mean one can not carry out Monte Carlo simulation on finite size system. If there is sign problem, the usual expedient \cite{Sugar1990exp} is to introduce a reference system which is sign-problem-free, by reweighting:
\begin{equation}
	\langle O\rangle=\frac{\sum_{\mathcal{C}} O(\mathcal{C}_i) W(\mathcal{C}_i)}{\sum_{\mathcal{C}} W(\mathcal{C}_i)}=\frac{\sum_{\mathcal{C}} O(\mathcal{C}_i) s(\mathcal{C}_i)\,|W(\mathcal{C}_i)| \,/ \sum_{\mathcal{C}}|W(\mathcal{C}_i)|}{\sum_{\mathcal{C}} s(\mathcal{C}_i)|W(\mathcal{C}_i)|\,/ \sum_{\mathcal{C}}\,|W(\mathcal{C}_i)|} \equiv \frac{\langle O s\rangle^{\prime}}{\langle s\rangle^{\prime}}
\end{equation}
where $s(\mathcal{C}_i)=\frac{W(\mathcal{C}_i)}{|W(\mathcal{C}_i)|}$ and $\langle O\rangle^{\prime}=\frac{\sum_{\mathcal{C}} O(\mathcal{C}_i) |W(\mathcal{C}_i)|}{\sum_{\mathcal{C}} |W(\mathcal{C}_i)|}$. The average of sign: $\langle s\rangle=Z / Z^{\prime}$, which is just the ratio of the partition function of the system $Z=\sum_{\mathcal{C}} W(\mathcal{C}_i)$ with weights $W(\mathcal{C}_i)$ and the reference system used for sampling with another partition function $Z^{\prime}=\sum_{\mathcal{C}}|W(\mathcal{C}_i)|$. 

As the partition function is usually an exponential form of the free energy, then $\langle s\rangle=Z / Z^{\prime}=\exp (-\beta N \Delta f)$, where $\Delta f$ is  the free energy densities differences of $Z$ and $Z^{\prime}$. When there is a sign problem, the term $\langle s\rangle^{\prime}$ in the denominator needs to be taken into account when calculating the mean value of quantity, so the relative error $\Delta s /\langle s\rangle$ of the sign are very important, and it usually increases exponentially with inverse temperature $\beta$ and system size $N$: 

\begin{equation}
	\frac{\Delta s}{\langle s\rangle}=\frac{\sqrt{\left(\left\langle s^{2}\right\rangle-\langle s\rangle^{2}\right) / N_{b}}}{\langle s\rangle}=\frac{\sqrt{1-\langle s\rangle^{2}}}{\sqrt{N_{b}}\langle s\rangle} \sim \frac{e^{\beta N \Delta f}}{\sqrt{N_{b}}}+O(\langle s\rangle)
\end{equation}
where $N_{b}$ is number of bins.  In most cases, $\Delta f$ is a quantity that doesn't change very much in the simulation, so when there a sign problem, the QMC simulation usually takes exponential time to have a reasonable estimation of the errorbars for physical observables.

The exponential computation time seems to imply that sign problem may be an nondeterministic polynomial (NP) problem, but is it an NP-hard problem? Here NP problem is the set of decision problems solvable in polynomial time by a nondeterministic Turing machine, and a NP-hard problem means every problem in NP can be reduced in polynomial time to it. At present, the answer still seems unclear, but explicit examples can deepen our understanding.  

People try to construct a simulation with sign problem, and the calculation of its physical quantity can be transformed into a solution of a NP-complete problem. Here a problem is NP-complete means it is both NP and NP-hard. It is suggested that if the polynomial time solution corresponding to the sign problem is found, the polynomial time solution of the NP-complete problem can also be found, that is, the sign problem is a NP-hard problem. In Ref.~\cite{NP-hard}, people consider a well-known special case of the classical Ising spin glass model. One can map it to NP-hard graph problem, such as 'finding ground energy' to another problem 'finding a cocycle of minimum weight while a two-level grid $G=(V, E)$, and a weighting function $J: E \rightarrow\{-1,0,1\}$ are given'~\cite{Barahona_1982}.  Here a graph $G=(V, E)$ consists of a set of vertices $V$, and a set $E$ of unordered pairs of different vertices, called edges. Then it is possible to introduce a corresponding NP-complete problem~\cite{NP-hard}: for given constant $C$, does the ground state have energy $E_0 \leq C$ ? And then one can map the classical Ising spin glass model to quantum Ising spin glass model with $\sigma^x_i \sigma^x_j$ term, which has sign problem. Mathematically, because this problem is NP-complete, there exists a polynomial time mapping to any other NP-complete problem.

However, the special spin glass example is equivalent to saying quantum spin glass problem is no easier to solve than the corresponding classical spin
glass problem. The example is very special and the proof does not address QMC solutions of systems, such as Hubbard model without particle-hole symmetry and SU(2) frustrated quantum spin models, which are not directly associated with the difficulties of simulating classical frustrated glassy systems with complicated energy landscapes. So we can not conclude that all sign problems are NP-hard. But from this example, at least we know that if we find a method to solve an NP-hard problem, this scheme can be used to solve QMC simulation with sign problem.

In addition, the definition $\langle s\rangle=Z / Z^{\prime}$ can give us some other information. Just write it in zero-temperatu limit:
\begin{equation}
\langle s \rangle_{R}=\frac{Z}{Z^{R}}=\frac{g e^{-\beta E_{0}}}{g^{R} e^{-\beta E_{0}^{R}}}
\label{eq:eq19}
\end{equation}
where $g$ and $g^{R}$ are the ground state degeneracy of systems $Z$ and refence system $Z^{R}$, while $E_0$ and $E^{R}_0$ are ground energy of systems $Z$ and $Z^{R}$. This means the expectation value $\langle s \rangle_{R}$ can be used to read $g$, $E_0$ for finite size systems, if we engineer the reference system with known $g_R$ and $E^{R}_0$. 

For spin system case, if the systems $Z$ and refence system $Z^{R}$ have same ground state, and the ground is in the set of QMC computational basis, one can see reduction of the sign problem at low-temperature, as shown in Refs.~\cite{SSmodel,SSmodel_basis}. For fermion case, we can also choose a good refence system $Z^R$, or choose a well-known $Z^{\prime} \geqslant Z^{R}$ to get the lower bound of the sign, this will be explained in the {\it Sign bound theory} in Sec.~\ref{sec:sec5.3}.

Now we know the definition of sign problem, then what is sign problem related to?  The collective efforts over the years have pointed out the sign problem have multiple origins. And we will discuss the major ones below.

\subsection{Sign Problem is basis-dependent}
First of all, the sign problem is basis-dependent. For example, if one can diagonalize the Hamiltonian matrix and calculate the eigenstates $\{ |i \rangle\}$ then $H|i\rangle=\varepsilon_{i}|i\rangle$ and
\begin{equation}
	\begin{aligned}
		\langle O\rangle & =\operatorname{Tr}[O \exp (-\beta H)] / \operatorname{Tr}[\exp (-\beta H)]     \\
		& =\sum_{i}\left\langle i\left|O_{i}\right| i\right\rangle \exp \left(-\beta \varepsilon_{i}\right) / \sum_{i} \exp \left(-\beta \varepsilon_{i}\right).
	\end{aligned}
\end{equation}
There's obviously no sign problem here. Unfortunately, it is often difficult to obtian eigenstates of a quantum many-body system in the first place. Moreover, if one already have the methods to obtain all eigenstates of the system, there is then no need to perform Monte Carlo simulations.   

In addition, for interacting fermion systems, different basis will bring different effects during decoupling. In a recent paper on automatic differentiation to mitigating the sign~\cite{wan2020mitigating}, the authors performed different decoupling at different sites $e^{-\Delta \tau U\left(\hat{n}_{i \uparrow}-\frac{1}{2}\right)\left(\hat{n}_{i \downarrow}-\frac{1}{2}\right)}=\frac{1}{2} e^{-U \Delta \tau / 4} \sum_{s_{i}=\pm 1} e^{\lambda s_{i} \hat{c}_{i}^{\dagger} \boldsymbol{\sigma} \cdot \boldsymbol{n}_{i} \hat{c}_{i}}$, where $\boldsymbol{n}_{i}=\left(\sin \theta_{i} \sin \phi_{i}, \sin \theta_{i} \cos \phi_{i}, \cos \theta_{i}\right)$. The choice of basis for decoupling at different sites can affect the severity of sign problem. While for spin systems, as discussed in Eq.~\eqref{eq:eq19}, if the ground state is a member of the computational basis, and sign-problem-free reference system have the same ground state, reduction of the sign problem at low-temperature~\cite{SSmodel,SSmodel_basis} will be seen.

We note there exist many research works that mitigate the sign problem by basis transformation~\cite{2015BasisChange,2021BasisChange2,2020BasisChange3,2020BasisChange4,2020BasisChange5,2019BasisChange6,2020BasisChange7,Rossi2017determinant,Rossi2017polynomial,DEmidio2020}. However, it has has also been pointed out that some sign problems turn out to be intrinsic~\cite{hastings2016quantum,ringel2017quantized,Intrinsic1,Intrinsic2} and cannot be eliminated by local unitary basis transformations.

\subsection{Sign Problem is related to Pauli exclusion principle}
\begin{figure}[!ht]
	\centering
	\includegraphics[width=0.4\linewidth]{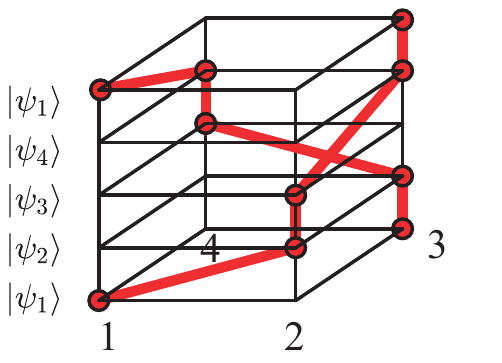}
	\caption{ An example of a configuration where particle exchange leads to negative weights. The dots represent particle occupation, and blod lines are world lines. }
	\label{fig:fig4}
\end{figure}

From the point of view of world-line Monte Carlo, the exchange of fermions in the world-line will bring a minus sign, which arise from the Pauli
exclusion principle. A simple such process is shown in Fig.~\ref{fig:fig4} and can be derived as
	\begin{equation}
	\begin{aligned}
	\langle \psi_1| \hat{c}_1^\dagger \hat{c}_4 \hat{c}_3^\dagger \hat{c}_2 \hat{c}_4^\dagger \hat{c}_3 \hat{c}_2^{\dagger}\hat{c}_1  | \psi_1\rangle&=\langle \psi_1|  \hat{c}_4 \hat{c}_3^\dagger \hat{c}_2 \hat{c}_4^\dagger \hat{c}_3 \hat{c}_2^{\dagger}\hat{n}_1  | \psi_1\rangle\\
	&=\langle \psi_1|  \hat{c}_4 \hat{c}_3^\dagger \hat{c}_4^\dagger \hat{c}_3 (1-\hat{n}_2)\hat{n}_1  | \psi_1\rangle\\
	&=-\langle \psi_1|  (1-\hat{n}_4)\hat{n}_3 (1-\hat{n}_2)\hat{n}_1  | \psi_1\rangle\\
	&=-\langle \psi_1 | \psi_1\rangle.
	\end{aligned}
	\end{equation}

\subsection{Sign Problem is related to geometric Frustration}

Similarly, frustrated spin or boson systems often have sign problem. For example, if we consider a boson (spin) system of three sites:
\begin{equation}
	\hat{H}=\left(\hat{b}_{1}^{\dagger} \hat{b}_{2}+\hat{b}_{2}^{\dagger} \hat{b}_{3}+\hat{b}_{3}^{\dagger} \hat{b}_{1}\right)+\text { h.c. \qquad or \qquad}  \hat{H}=\left(\hat{S}_{1}^{+} \hat{S}^-_{2}+\hat{S}_{2}^{+} \hat{S}^{-}_{3}+\hat{S}_{3}^{+} \hat{S}^{-}_{1}\right)+\text { h.c.} 
\end{equation}
The weight is negative for the following configuration (where $|i\rangle$ means i-th site occupied state):
\begin{equation}
	\left\langle 1\left|e^{-d \beta \hat{H}}\right| 2\right\rangle\left\langle 2\left|e^{-d \beta \hat{H}}\right| 3\right\rangle\left\langle 3\left|e^{-d \beta \hat{H}}\right| 1\right\rangle, \text { contribution }=(-1)^{3} d \beta^{3}.
\end{equation}

However, frustration does not necessarily mean there must be a sign problem. Clever choice of basis and make the QMC simulation sign-problem free. As discussed in Sec.~\ref{sec:sec3.2}, there is no sign problem for some bilayer frustrated models~\cite{Frustrated_bilayer,Frustrated_dimer,Frustrated_ladders,bilayer_again,bilayer_b}. Inspired by these works and considering the model that has the same ground state as the above model, one can extend the simulation basis from spin to singlet-triplet or even plaquette bases, and recently there are progresses in such cases by means of changing the simulation basis from spin to bond or even plaquette~\cite{SSmodel,SSmodel_basis} to reduce the sign problem in the QMC simulation for 2D antiferromagnetic Shastry-Sutherland spin model.

\subsection{Sign Problem is related to Aharonov-Anandan Phase}

For DQMC, the sign problem is thought to be related to topological properties of the configurations~\cite{Topological_Origin}. One can define the weight as:
\begin{equation}
	W(\mathcal{C}_i)=\operatorname{det}\left[1+e^{\beta \mu} \mathcal{T} \exp \int_{0}^{\beta} \mathrm{d} \tau H^{\operatorname{aux}}[\mathcal{C}_i(\tau)]\right],
\end{equation}
then define
\begin{equation}
	G(\tau ; \beta)=\mathcal{T} \exp \int_{\tau}^{\tau+\beta} \mathrm{d} \tau^{\prime} H^{\text {aux }}\left[\mathcal{C}_i\left(\tau^{\prime}\right)\right],
\end{equation}
the corresponding eigenvalue and eigenstate are: $G(\tau ; \beta)\left|\phi_{n}(\tau)\right\rangle=\lambda_{n}\left|\phi_{n}(\tau)\right\rangle $.

Now one finds the weight is 
\begin{equation}
	W(\mathcal{C}_i)=\prod_{n}\left(1+\lambda_{n} e^{\beta \mu}\right)
\end{equation}
by defining $\lambda_{n}=e^{i \theta_{n}} \omega_{n}$ and $e^{i \theta_{n}}=\prod_{\tau=0}^{\beta}\left\langle\phi_{n}(\tau+d \tau) \mid \phi_{n}(\tau)\right\rangle$. These phases $e^{i \theta_{n}}$ are imaginary time versions of the Aharonov-Anandan (AA) phase. If we take $\beta \rightarrow \infty$, which means the adiabatic limit, then the eigenstate $\left|\psi_{n}(\tau)\right\rangle=\lim_{\beta \rightarrow \infty}\left|\phi_{n}(\tau)\right\rangle$ and the weights is $\exp \left[-\int_{0}^{\beta} \varepsilon_{n}(\tau) d \tau\right]$, where $\varepsilon_{n}(\tau)$ is eigenvalue of $H^{\text {aux }}(\tau)$. Now $e^{i \theta_{n}}$ becames a Berry phase of the instantaneous eigenstates $\left|\phi_{n}(\tau)\right\rangle$.

This analogy gives us a new perspective on the sign problem, but it doesn't help to mitigate it. In the next Section, we will discuss understanding and practical implementation scheme to cure, ease and make use of the sign problem.  

\section{How to cure, ease and make use of the sign problem}
\label{sec:sec5}
How to reduce the exponential computation complexity introduced by the sign problem, there is no general methods, and there are different solutions for different physical systems. We note only the recent {\it Sign bound theory} has pointed out a generic direction~\cite{zhang2021sign}. Below, we select few important cases and explain their usages.

\subsection{Symmetry}
In DQMC for interacting fermion models, where $H=H_{K}+H_{I}$, $H_{K}$ is kinetic energy part and $H_{I}$ is the part of the interaction after decoupling~\cite{2005CongjunWu}, if there exists an anti-unitary operator $T$, such that
\begin{equation}
	T H_{K} T^{-1}=H_{K}, \quad T H_{I} T^{-1}=H_{l}, \quad T^{2}=-1
\end{equation}
then for the eigenvalues $\lambda_i$ of matrix $I+B(\beta,0)$, where $B(\beta,0)$ is defined in Eq.~\eqref{eq:eq15}. We know $\prod \lambda_i$ is real and positive number.

The proof is simple. If $(I+B)\left|\Psi_{i}\right\rangle=\lambda_{i}\left|\Psi_{i}\right\rangle$ then $T\left|\Psi_{i}\right\rangle$ is also eigenstate of matrix $(I+B)$ with eigenvalue $\lambda_{i}^{*} $. That is, the eigenvalues of $\lambda_{i} $ and $\lambda_{i}^{*} $ come in pairs, and the weight $\det (I+B)=\prod |\lambda_i|^2 \geqslant 0$. The orthogonality of the two eigenstates is trivial: $\langle \Psi_{i}\left | T \Psi_{i}\right\rangle=\langle \Psi_{i}|T^{\dagger}TT\left|\Psi_{i}\right\rangle^{*}=-\langle \Psi_{i}|T^{\dagger}\left|\Psi_{i}\right\rangle ^{*} = - \langle T \Psi_{i}\left|\Psi_{i}\right\rangle ^{*} =-\langle \Psi_{i}\left|T \Psi_{i}\right\rangle=0 $. The overlap of the two states is zero.

In practice, one usually divides the configurational weight of the original problem into two parts according to the degrees of freedom, such as the spin up and down, then perform a unitary transformation to show the two parts of the weights are either the same real number or complex conjugate with each other, such as their products is positive definite. For example, Hubbard Model at half filling, if $U>0$, $e^{-\Delta \tau \hat{H}_{U}}=\gamma \sum_{s=\pm 1} e^{\alpha s\left(\hat{n}_{\uparrow}+\hat{n}_{\downarrow}-1\right)}$ where $\gamma=\frac{1}{2} e^{\Delta \tau U / 4}, \; \cosh (\alpha)=e^{-\Delta \tau U / 2}$, and $\alpha$ is an imaginary number. Now the unitary transformation is the particle-hole transformation, 
\begin{equation}
	\left\{\begin{array}{l}	
		\hat{c}_{i \downarrow}^{\dagger} \rightarrow(-1)^{i} \hat{c}_{i \downarrow} \\	
		\hat{c}_{i \downarrow} \rightarrow(-1)^{i} \hat{c}_{i \downarrow}^{\dagger}	
	\end{array}\right.
\end{equation}
then\begin{equation}
	\text {$\alpha$ is an imaginary number} \rightarrow\left\{\begin{array}{c}	
		\alpha\left(\hat{n}_{i \uparrow}+\hat{n}_{i \downarrow}-1\right)\stackrel{P H \text { for spin-down }}{\longrightarrow} \alpha s_{i}\left(\hat{n}_{i \uparrow}-\hat{n}_{i \downarrow}\right) \\	
		e^{\alpha s_{i}\left(\hat{n}_{i \uparrow}+\hat{n}_{i \downarrow}-1\right)} \stackrel{P H \text { for spin-down }}{\longrightarrow} e^{\alpha s_{i}\left(\hat{n}_{i \uparrow}-\hat{n}_{i \downarrow}\right)}	
	\end{array}\right.	
\end{equation}
Since $\alpha$ is an imaginary number, it's clear to see that after such a particle-hole transformation, spin up and down parts are complex conjugated to each other.
 
In addition to particle-hole symmetry, other symmetries such as $C_2 T$ + $C_2 P$ symmetries \cite{XuZhang2021,JYLee2021} may also make the system sign-free. In short, we want to find an anti-unitary transformation that makes the system invariant, or to apply a unitary transformation, then the weights are divided into two parts that are obviously complex conjugate to each other. Such scheme has been widely used in the DQMC solution of interaction driven topological phase transitions, such as in Kane-Mele Hubbard model~\cite{Hohenadler2012,meng2014}.
  	
\subsection{Split Orthogonal Group, Majorana representation and Majorana Positivity}

Subsequently, people have made more in-depth studies on sign of determinants of matrices. For example, there is a theorem\cite{2015LeiWang} :  

If $M$ belongs to the split orthogonal group $O(n, n)$, then  :
\begin{equation}
	\operatorname{det}(I+M) \begin{cases}\geqslant 0, & \text { if } M \in O^{++}(n, n), \\ \leq 0, & \text { if } M \in O^{--}(n, n), \\ 0, & \text { otherwise. }\end{cases}
\end{equation}

where the split orthogonal group $O(n, n)$ is formed by all $2 n \times 2 n$ real matrices that preserve the metric $\eta=\operatorname{diag}(\underbrace{1, \ldots,}_{n}, \underbrace{-1, \ldots,-1}_{n})$, $M^{T} \eta M=\eta$. Split orthogonal group $O(n, n)$ means that if we write the matrix $M$ in the block form like:$M=\left(\begin{array}{ll}M_{11} & M_{12} \\ M_{21} & M_{22}\end{array}\right)$, then we have $\left|\operatorname{det}\left(M_{11}\right)\right| \geqslant 1$ and $\left|\operatorname{det}\left(M_{22}\right)\right| \geqslant 1$. And we use the notation $O^{\pm \pm}(n, n)$ to classified the four components of $O(n, n)$ by considering the signs of $\operatorname{det}\left(M_{11}\right)$ and $\operatorname{det}\left(M_{22}\right)$.

Since the fermion sign problem is related to the basis of decoupling, it can be found that some models translated to Majorana representation are sign-problem free~\cite{HongYao1,HongYao2,HongYao3}. Majorana representation is turning fermions into Majorana fermions:
\begin{equation}
	\hat{c}_{i}=\frac{1}{2}\left(\gamma_{i}^{1}+i \gamma_{i}^{2}\right), \quad \hat{c}_{i}^{\dagger}=\frac{1}{2}\left(\gamma_{i}^{1}-i \gamma_{i}^{2}\right)
\end{equation}
These findings inspired research on Majorana reflection positivity or the Majorana Kramers positivity~\cite{2016TaoXiang}. One can write the $H_0$ and decoupled Hamiltonian under the Majorana representation: $H_{\mathrm{bl}}=\gamma^{T} V \gamma$. Then 

(a) if $V$ is  a Majorana reflection positive kernel, which means it can be represented as
\begin{equation}
	V=\left(\begin{array}{cc}
		A & i B \\
		-i B^{T} & A^{*}
	\end{array}\right)
\end{equation}
where $A$ and $B$ are $N \times N$ matrices. $A$ is complex antisymmetric matrix and $B$ is a Hermitian matrix. Or,

(b) if there exist two transformation operators $S$ and $P$ such that
\begin{equation}
	\begin{aligned}
		&S^{T} V S=V^{*}, \\
		&P V P^{-1}=V,
	\end{aligned}
\end{equation}
where $S$ is a real antisymmetric matrix and $P$ is a symmetric or antisymmetric Hermitian matrix, and we need $P$ anticommutes with $S$. There is no sign problem in the DQMC simulation.

In Ref.~\cite{XuMonteCarlo2019}, there is a similar discussion on the sign problem related with the structure of the fermion determint. It is proved that if $D \in \mathrm{SU}(n, m)$, then $\operatorname{det}\left(I_{n+m}+D\right) \in \mathbb{R}$, where $ \mathrm{SU}(n, m)$ is a pseudounitary group. Such condition is used to guarantee the DQMC simulation for U(1) gauge field coupled to Dirac fermions in $(2+1)$d, a fundamental problem for both high-energy physics and condensed matter physics related to quantum electrodynamics and deconfined quantum criticalities, is sign-problem free~\cite{XuMonteCarlo2019,weiwang2019,Janssen2020}.

\subsection{Sign bound theory}
\label{sec:sec5.3}
As discussed in Sec.~\ref{sec:sec4}, for systems with sign problems, we can usually sample them by reweighting and the average sign is usually $ \langle s\rangle=Z/Z^{\prime}\sim e^{-\beta N \Delta f}$.   However, the difference of free energy $\Delta f$ between the reference system $Z^{\prime}$ and the system $Z$ is not necessarily a constant with little change, but may be a quantity depending on the system size $N=L^d$ and inverse temperature $\beta=\frac{1}{k_B T}$. 

In some systems, we find that in the zero temperature limit, the sign does not decay exponentially with respect to size $L$ but rather decay algebraically~\cite{ouyang2021projection,zhang2021sign}, impling that $\Delta f$  could be a logarithmic function of $L$.

In fact, in the zero temperature limit, the mean of the signs is
$\langle s \rangle_{R}=\frac{Z}{Z^{R}}=\frac{g e^{-\beta E_{0}}}{g^{R} e^{-\beta E_{0}^{R}}}$, as given in Eq.~\eqref{eq:eq19}. We can choose a good refence system $Z^{R}$, which have same ground state energy $E^{R}_0=E_0$ with $Z$. Then the mean of the signs is
$\langle s \rangle_{R}=g/g^R$, which is only related to the ratio of ground state degeneracy. And as shown in the cases below, in a few common cases, such as the lattice model for quantum Moir\'e materials in momentum space~\cite{XuZhang2021,GaopeiPanValley2021,zhang2021sign,XuZhang2021SC} and extended Hubbard model in real space~\cite{ouyang2021projection,LiaoCPB2021,YDLiaoPRX2021,YDLiao2022}, the ground state degeneracy is a polynomial function of the size of the system.

 Even if most of the time we can not find such a perfect $Z^R$, we can still consider the upper bound on the value of $Z^R$, which means if we know $Z^R \leq C $, then in the zero temperature limit we can get a lower bound of the sign: $\langle s \rangle_{R}=\frac{Z}{Z^{R}}\geqslant \frac{g e^{-\beta E_{0}}}{C}$. This is the {\it Sign bound theory}~\cite{zhang2021sign}. Two cases of the theory will be shown below. In case 1: $Z^R =g^{R} e^{-\beta E_{0}^{R}} \leq \sqrt{g^{\prime} e^{-\beta E_{0}^{\prime}} } $ and $E_0^\prime/2=E_0$ such that $\langle s \rangle_{R}=\frac{Z}{Z^{R}}\geqslant \frac{g e^{-\beta E_{0}}}{\sqrt{g^{\prime} e^{-\beta E_{0}^{\prime}} }}=g/\sqrt{g^\prime} e^{-\beta (E_{0}-E_0^\prime/2)}=g/\sqrt{g^\prime}$, and in case 2: $Z^R =g^{R} e^{-\beta E_{0}^{R}} \leq g^{\prime} e^{-\beta E_{0}^{\prime}} $ and $E_0^{\prime}=E_0$ such that $\langle s \rangle_{R}=\frac{Z}{Z^{R}}\geqslant \frac{g e^{-\beta E_{0}}}{g^{\prime} e^{-\beta E_{0}^{\prime}}}=g/g^\prime e^{-\beta (E_{0}-E_0^\prime)}=g/g^\prime$ . If we succeed in finding such an upper bound on $Z^{R}$ which can cancel out $e^{-\beta E_{0}}$ term in the denominator , then the lower bound of the sign is just a function of the ground state degeneracy. If the ground state degeneracy is a polynomial function of the size of the system, then we can obtain a lower bound of the averge sign algebraically decay with the system size.

\subsubsection{Case 1}
For example, we choose $Z^{R}=\sum_{\{l\}} P(\{l\}) |\mathcal{R}(D(\{l\}))| \equiv \langle |\mathcal{R}(D)|\rangle$ and $Z^{\prime}=\sum_{\{l\}} P(\{l\}) |D|^2(\{l\}) \equiv\langle |D|^2\rangle$, where $Z=\sum_{\{l\}} P(\{l\}) D(\{l\}) \equiv\langle D\rangle = \langle \mathcal{R}(D)\rangle$. Here $ D(\{l\})$ is the determinant term in weight, and $P(\{l\})$ is a normalized probability distribution. Because $\langle |\mathcal{R}(D)|\rangle\leq \sqrt{\langle |D|^2\rangle}$  then $\langle s \rangle=\frac{Z}{Z^{R}}\geqslant \frac{g e^{-\beta E_{0}}}{\sqrt{g^{\prime} e^{-\beta E_{0}^{\prime}} }}=g/\sqrt{g^\prime} e^{-\beta (E_{0}-E_0^\prime/2)}=g/\sqrt{g^\prime}$, if $E_0^{\prime}/2=E_0$.
	
The twisted-bilayer-graphene(TBG) models, projected to the flat bands, belong to such a case~\cite{XuZhang2021,GaopeiPanValley2021}. In the chiral limit, the Hamiltonian only has density-density interaction in momentum space
\begin{equation}
 H=\sum_{q \neq 0} V(q) \rho_{-q} \rho_{q}=\sum_{q \neq 0} V(q) \rho_{q}^{\dagger} \rho_{q}
\end{equation}
 where $\rho_{q}=\sum_{i, j}\left(\lambda_{i, j}(q) \hat{c}_{i}^{\dagger} c_{j}-\frac{1}{2} \mu_{q}\right)$. 

\begin{figure}[!htp]
 	\includegraphics[width=1.0\textwidth]{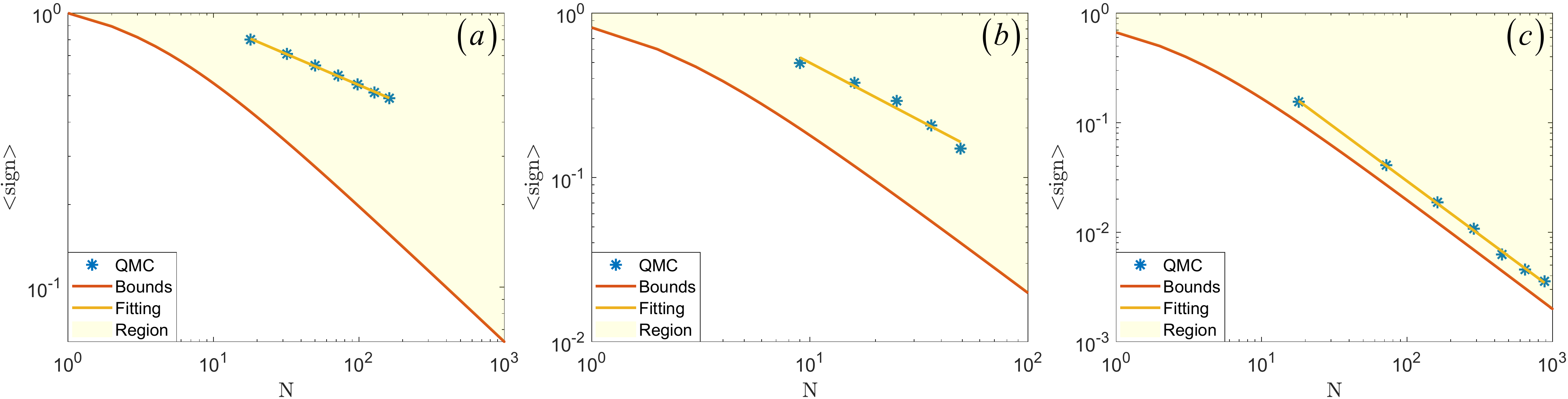}
 	\caption{The average sign for three different cases according to the {\it sign bound theory}. All measurement are carried out at a average sign converged temperature. The errorbars in QMC data are smaller than the symbol size.  (a) momentum space case, now $\lambda$ is randomly set, and the red line gives us the lower bound $\left\langle s\right\rangle \geqslant  {2}/{\sqrt{N+3}}$. Here $N=18,32,50,72,98,128,162$. Fitting line $\propto$ $N^{-0.23}$. (b) momentum space case, we consider $\nu=0$ single valley/spin TBG model in chiral limit, and the red line gives us the lower bound $\left\langle s\right\rangle \geqslant  2 / {\sqrt{(N+1)^2+2}}$. $N=9,16,25,36,49$ and $T=0.91$ meV. Fitting line $\propto N^{-0.70}$. (c) For real space TBG model, the red line lower bound is $\left\langle s\right\rangle \geqslant  2/(N+2)$ and $N=18,72,162,288,450,648,882$. The parameters of QMC are detailed in ~\cite{ouyang2021projection}. Fitting line $\propto N^{-0.98}$.}
 	\label{fig:fig5}
 \end{figure}

If we set $\lambda$ is random number and $\mu_q=0$, and there is no other symmetry, then $E_0=E_0^\prime/2$. For ground state degeneracy, it's easy to know $g=2$ and $g^{\prime}=N+3$, by introducing a raising operator
 $\Delta^{\dagger}=\sum_{i^{\prime}} c_{i^{\prime},+}^{\dagger} c_{i^{\prime},-}$. Then at low temperature,$\langle s\rangle \geqslant  2 / \sqrt{N+3}$ (as shown in Fig.~\ref{fig:fig5} (a)).

We consider single valley/spin two band ($m, n \in\{1,-1\}$) TBG model at hall-fing and chiral limit, which means $\lambda_{i, j, m, n}(q)=m \cdot n \cdot \lambda_{i, j, m, n}(q)$ and $\mu_q \neq 0$ . Similarly, we still have $E_0=E_0^\prime$ and $g=2$, and now $g^{\prime}=(N+1)^{2}+2$. Then at low temperature,$\langle s\rangle \geqslant  2 / \sqrt{(N+1)^{2}+2}$ (as shown in Fig.~\ref{fig:fig5} (b)).

 \subsubsection{Case 2}
If we consider the real space extended Hubbard model~\cite{ouyang2021projection,LiaoCPB2021,YDLiaoPRX2021}, which also offer the description of the TBG systems once the interactions are projected to the flat bands with truncation:
\begin{equation}
\hat{H}=U \sum_{\varhexagon}\left(\hat{Q}_{\varhexagon}+\alpha \hat{T}_{\varhexagon}-\nu\right)^{2}
\end{equation}
where $\hat{Q}_{\varhexagon}=\frac{1}{3} \sum_{\sigma, \tau} \sum_{l=1}^{6} \hat{c}_{R+\delta_{l}, \sigma, \tau}^{\dagger} \hat{c}_{R+\delta_{l}, \sigma, \tau}-4, \hat{T}_{\varhexagon}=$ $\sum_{\sigma, \tau} \sum_{l=1}^{6}\left[(-1)^{l} \hat{c}_{R+\delta_{l+1}, \sigma, \tau}^{\dagger} \hat{c}_{R+\delta_{l}, \sigma, \tau}+\right.$ h.c. $], \nu$ is used to control filling, $\sigma, \tau$ are spin and valley indexes, $R+\delta_{l}$ represents site $l$ in a single $R$ hexagon and $U, \alpha$ is real number coming from the overlap of Wannier functions on the Moir\'e scale. 

Here, we choose $Z^{\prime}=\sum_{\{l\}} P(\{l\}) |D|(\{l\}) \equiv\langle |D|\rangle$, still $Z^{R}=\sum_{\{l\}} P(\{l\}) |\mathcal{R}(D(\{l\}))| \equiv \langle |\mathcal{R}(D)|\rangle$ and $Z=\sum_{\{l\}} P(\{l\}) D(\{l\}) \equiv\langle D\rangle = \langle \mathcal{R}(D)\rangle$. Because $ \langle |\mathcal{R}(D)|\rangle \leq \langle |D|\rangle$  then $\langle s \rangle=\frac{Z}{Z^{R}}\geqslant \frac{g e^{-\beta E_{0}}}{g^{\prime} e^{-\beta E_{0}^{\prime}}}=g/g^\prime e^{-\beta (E_{0}-E_0^\prime)}=g/g^\prime$. Similarly, we still have $E_0^\prime=E_0$ . It can be computed from tensor Young tableau method that, at $\nu=\pm 2$ , $g=(N+3)(N+2)(N+1) / 6$, and at $\nu=0$, $g^{\prime}=(N+3)(N+2)^{2}(N+1) / 12$~\cite{zhang2021sign}. Then at low temperature,$\langle s\rangle \geqslant  g / g^\prime=2 /(N+2)$ (as shown in Fig.~\ref{fig:fig5} (c)).

For the model we mentioned above and its reference system, the exponential decay part cancels out, and we can easily calculate the ground state degeneracy. As shown in Fig.~\ref{fig:fig5}, numerically we also see that at very low temperatures, the average sign is polynomial decay.

\subsection{Lefschetz thimble}
This is also a method that can be used to cure or reduce the sign problem in lattice fermion QMC simulations~\cite{Lef2,Lef3,cristoforetti2013monte,fukuma2019applying,mukherjee2014lefschetz,mukherjee2013metropolis,mooney2021lefschetz,PhysRevD.104.074517,Lef11}, here we only briefly outline its main idea. 

For computing expectation value $\langle \mathcal{O}[x] \rangle$ in Monte Carlo sampling on the configuration space $[x]$, we have
\begin{equation}
	\begin{aligned}
		\langle\mathcal{O}[x]\rangle & =\frac{1}{Z} \int d x e^{-S[x]} \mathcal{O}[x]   \\
		& =\frac{\int d x e^{-S_{R}[x]} e^{-i S_{I}[x]} \mathcal{O}[x]}{\int d x e^{-S_{R}[x]} e^{-i S_{I}[x]}}   \\
		& =\frac{\int d x e^{-S_{R}[x]} e^{-i S_{I}[x]} \mathcal{O}[x]}{\int d x e^{-S_{R}[x]}} / \frac{\int d x e^{-S_{R}[x]} e^{-i S_{I}[x]}}{\int d x e^{-S_{R}[x]}} \\
		& =\frac{\left\langle e^{-i S_{I}} \mathcal{O}\right\rangle_{R}}{\left\langle e^{-i S_{I}}\right\rangle_{R}}
	\end{aligned}
	\label{eq:eq36}
\end{equation}
where $R$ and $I$ means real and imaginary parts of action $S[x]$. And $\left\langle \; \right\rangle_{R}$ is defined as $\left\langle O\right\rangle_{R}=\frac{\int d x e^{-S_{R}[x]} O}{\int d x e^{-S_{R}[x]}}$.
 
If action $S[x]$ is analytic for all $x \in \mathbb{C}^{n}$, its saddle points are nondegenerate and Eq.~\eqref{eq:eq36} is convergent, then Morse theory~\cite{banyaga2004lectures,witten2010new,witten2011analytic} told us these integrals can be evaluated using the steepest descent cycles $\mathcal{J}_{\sigma}$ and they are called Lefschetz thimbles, which means we can replace the
integration over the real domain $R^n$ with that over curved complex n-dimensional manifolds $\mathcal{J}_{\sigma}$
\begin{equation}
	\int_{\mathbb{R}^{n}} \mathrm{~d}^{n} x e^{- S(x)}=\sum_{\sigma} n_{\sigma} \int_{\mathcal{J}_{\sigma}} \mathrm{d}^{n} z e^{- S(z)}
\end{equation}
where the Lefschetz thimbles $\mathcal{J}_{\sigma}$ are the union of all solutions of the gradient flow equations:
\begin{equation}
\frac{d z(\tau)}{d \tau}=-\frac{\overline{\partial S}}{\partial z}
\end{equation} 
 and it start from  the corresponding saddle point $z_{\sigma}$, which means $\lim_{\tau \rightarrow  \infty} z(\tau)= z_{\sigma}$ . Here $\sigma$ is the index for different saddle point/Lefschetz thimbles, $\tau$ is a parameter and the overline represents complex conjugation, and saddle point is the point on the surface of $S$ where the derivatives in orthogonal directions are all zero, but is not a local extremum. While the thimbles always end inside regions of stability, duals thimbles(anti-thimbles) $\mathcal{K}_{\sigma}$ are also such solutions at a given saddle point $\lim_{\tau \rightarrow  -\infty} z(\tau)= z_{\sigma}$ and end inside regions of instability. We can make an analogy between the dual thimble and the contour of steepest ascent, as we did with thimble and the contour of steepest descent. And $n_{\sigma}$ are intersection numbers (which decides the contribution of a particular saddle point $z_{\sigma}$ to the partition function) of hypersurface generated by the paths of steepest ascent duals $\mathcal{K}_{\sigma}$ and original region of integration $\mathbb{R}^{n}$ :
 \begin{equation}
n_{\sigma}=\left\langle\mathcal{K}_{\sigma},\mathbb{R}^{n}\right\rangle
 \end{equation}
 
The important point of such construction is: $\operatorname{Im}[S]$ is constant on Lefschetz thimbles (mod $2\pi$), which means there is no sign problem on each thimble. The ratio of weights in the same thimble must be a positive real number, since the phase of the weight is the same.

\begin{figure}[!htp]
 	\centering
 	\includegraphics[width=0.4\textwidth]{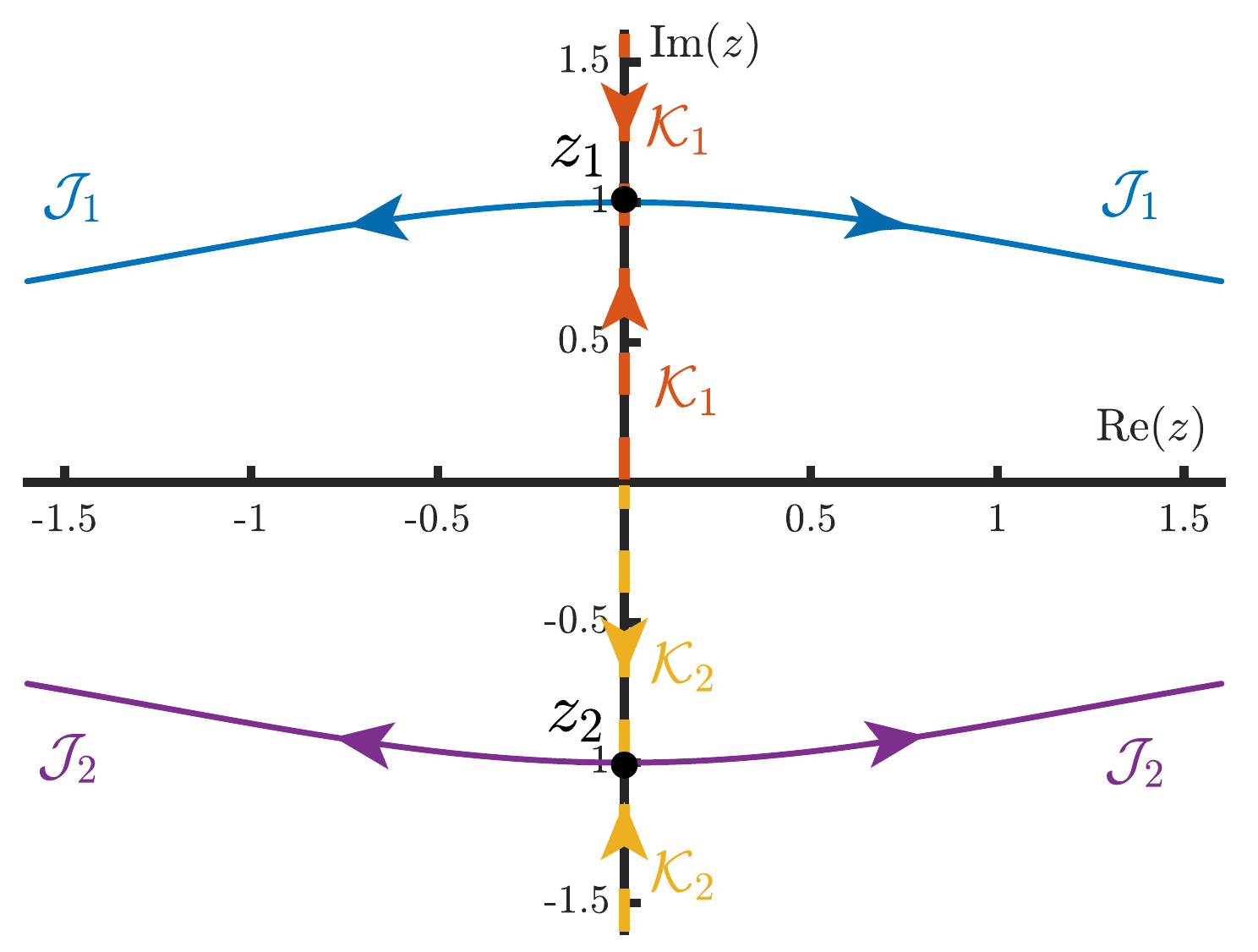}
 	\caption{A schematic plot of thimbles and dual thimbles. Here the black dots are saddle points $z_{1,2}$, while the solid blue/purple lines correspond to thimbles and the dotted red/yellow lines correspond to dual thimbles. And the arrows show the directions of the flows. See intersection numbers of dual thimbles and $\mathbb{R}+i \varepsilon$ , we know $n_1=1,\;n_2=0$.}
 	\label{fig:fig6}
 \end{figure}

Let's take a trivial 1d case for the purpose of illustration\cite{kanazawa2015structure,bharathkumar2020lefschetz}:
\begin{equation}
\int_{\mathbb{R}+i \varepsilon} \mathrm{d} x \;\frac{1}{x} \mathrm{e}^{-x^{2} / 2}=\sum_{\sigma} n_{\sigma} \int_{\mathcal{J}_{\sigma}} \mathrm{d} z\; e^{- S(z)}
\end{equation}
which corresponds to the case $S[z]=\log z+z^{2} / 2$. The two critical points of this action are: $z_1= i,\,z_2= -i$. Since we know that the imaginary parts $\operatorname{Im}[S]$ are constant on Lefschetz thimbles, then we have: $\left.\operatorname{Im}(S[z])\right|_{\text {along } \mathbb{R}+i \varepsilon}=\operatorname{Im} S\left[z_{\sigma}\right]=\pm i \pi/2$. Following the procedure described above we can obtain the corresponding thimbles and dual thimbles(as shown in Fig.~\ref{fig:fig6}), and then: $
\int_{\mathbb{R}+i \varepsilon} \mathrm{d} x \;\frac{1}{x} \mathrm{e}^{-x^{2} / 2}= \int_{\mathcal{J}_{1}} \mathrm{d} z\; e^{- S(z)}$.
 
The Lefschetz thimbles method is commonly used to calculate the value of certain integrals and in high-energy physics and real time calculations. Recently with the application of hybrid Monte Carlo in condensed matter, there has been a lot of work using Lefschetz thimbles method to study common condensed matter problems which has sign problem. For example, Hubbard model on the hexagonal(honeycomb) lattice at finite chemical potential with linear system size $L=6$ and inverse tempearture $\beta=20$, has been successfully simulated with Lefschetz thimbles method in DQMC~\cite{Lef2,Lef11}.

\section{Conclusion}
In this article, we give a pedagogical overview on the origin of the sign problem in various quantum Monte Carlo simulation techniques, ranging from the world-line and stochastic series expansion Monte Carlo for boson and spin systems to the determinant and momentum space quantum Monte Carlo for interacting fermions. We have elaborated on the definition of the sign problem and its possible origins, including Pauli exclusion principle, geometric frustration and the lack of symmetry requirements, etc. In addition, we also point out the sign problem is actually basis-dependent and summarizes established ways to ensure that some models are free of sign problems such as checking symmetry and Majorana positivity. We explain what to do when there is sign problem in general: reweighting, and how to reduce the severity of sign problems, in particular that based on the properties of the finite size partition functions, the recent {\it{sign bound theory}} could distinguish when the bounds have the usual exponential scaling, and when they are bestowed with an algebraic scaling at low temperature limit. Fermionic QMC simulations with such algebraic sign problems have been successfully carried out for extended Hubbard-type and quantum Moir\'e lattice models.

\section*{Acknowledgement} We thank Xu Zhang, Weilun Jiang, Yuan Da Liao and Xiao Yan Xu for insightful discussions and fruitful collaborations on related topics over the years. GPP and ZYM acknowledge the support from the Research Grants Council of Hong Kong SAR of China (Grant Nos.~17303019, 17301420, 17301721 and AoE/P-701/20), the Strategic Priority Research Program of the Chinese Academy of Sciences (Grant No. XDB33000000), the K. C. Wong Education Foundation (Grant No. GJTD-2020-01) and the Seed Funding “QuantumInspired explainable-AI” at the HKU-TCL Joint Research Centre for Artificial Intelligence. We thank the Information Technology Services at the University of Hong Kong and the Tianhe platforms at the National Supercomputer Centers for their technical support and generous allocation of CPU time.

\bibliographystyle{unsrt}
\bibliography{Sign.bib}

\end{document}